\documentclass[aps,prd,amsmath,superscriptaddress,nofootinbib,twocolumn]{revtex4}
\pdfoutput=1

\usepackage{graphicx}
\usepackage{hyperref}

\newcommand{\ie}{\textit{i.e.}}
\newcommand{\eg}{\textit{e.g.}}

\newcommand{\rmd}{\ensuremath{\mathrm{d}}}
\newcommand{\orderof}[1]{\ensuremath{\mathcal{O}(#1)}}

\newcommand{\gae}{%
  \ensuremath{\lower 2pt \hbox{%
    $\, \buildrel {\scriptstyle >}\over {\scriptstyle \sim}\,$}%
    }%
  }

\newcommand{\lae}{%
  \ensuremath{\lower 2pt \hbox{%
    $\, \buildrel {\scriptstyle <}\over {\scriptstyle \sim}\,$}%
    }%
  }

\newcommand{\mpl}{\ensuremath{m_\textrm{Pl}}}
\newcommand{\mgut}{\ensuremath{m_\textrm{GUT}}}
\newcommand{\phie}{\ensuremath{\phi_\textrm{e}}}
\newcommand{\Ve}{\ensuremath{V_\textrm{e}}}

\newcommand{\ns}{\ensuremath{n_s}}
\newcommand{\nt}{\ensuremath{n_\textrm{T}}}

\newcommand{\Pscalar}{\ensuremath{P_\mathcal{R}}}
\newcommand{\Pscalarrt}{\ensuremath{\Pscalar^{1/2}}}
\newcommand{\Ptensor}{\ensuremath{P_T}}
\newcommand{\Ptensorrt}{\ensuremath{\Ptensor^{1/2}}}
\newcommand{\Pratio}{\ensuremath{\frac{P_T}{P_\mathcal{R}}}}
\newcommand{\rhoRH}{\ensuremath{\rho_\textrm{RH}}}

\newcommand{\refeqn}[2][eqn:]{Eq.~(\ref{#1#2})}

\newcommand{\reffig}[2][fig:]{Figure~\ref{#1#2}}

\newcommand{\refsec}[2][sec:]{Section~\ref{#1#2}} 

\newcommand{\insertfig}[1]{%
    \includegraphics[keepaspectratio,width=1.00\columnwidth,
                     height=0.40\textheight]{#1}
}

\begin{document}

\preprint{MCTP-06-23}
\preprint{FTPI-MINN-06/32}

\title{Natural Inflation: Consistency with Cosmic Microwave Background Observations of Planck and BICEP2}

\author{Katherine Freese}
\email[]{ktfreese@umich.edu}
\affiliation{
 Department of Physics,
 University of Michigan,
 Ann Arbor, MI 48109}

\author{William H. Kinney}
\email[]{whkinney@buffalo.edu}
\affiliation{
 Department of Physics,
 University at Buffalo, SUNY,
 Buffalo, NY 14260}

\date{\today}

\begin{abstract}

 Natural inflation is a  good
fit to all cosmic microwave background (CMB) data and may be the correct description of an early inflationary expansion of the Universe.
The large angular scale CMB polarization experiment BICEP2 has announced a major discovery,  which can be explained as
 the gravitational wave signature of inflation, at a level that matches predictions by natural inflation models.
 The natural inflation (NI) potential is theoretically exceptionally well motivated in that it is naturally flat due to
shift symmetries, and in the simplest version takes the form $V(\phi)
= \Lambda^4 [1 \pm \cos(N\phi/f)]$.  A tensor-to-scalar ratio $r>0.1 $ as seen by BICEP2 requires the height of any inflationary
potential to be comparable to the scale of grand unification and the
width  to be comparable to the Planck scale.   The Cosine Natural Inflation model agrees with
all cosmic microwave background measurements as long as $f \gae  \mpl$ (where
$\mpl = 1.22 \times 10^{19}\ {\rm GeV}$) and $\Lambda \sim \mgut \sim 10^{16}\ {\rm GeV}$.    
This paper also discusses other 
 variants of the natural inflation paradigm: we show that axion monodromy with potential $V\propto \phi^{2/3}$ 
 is inconsistent with the BICEP2 limits at the 95\% confidence level, and low-scale inflation is strongly ruled out. Linear potentials $V \propto \phi$ are inconsistent with the BICEP2 limit at the 95\% confidence level, but are marginally consistent with a joint Planck/BICEP2 limit at 95\%. We discuss the pseudo-Nambu Goldstone model proposed by Kinney and Mahanthappa as a concrete realization of low-scale inflation. While the low-scale limit of the model is inconsistent with the data, the large-field limit of the model is marginally consistent with BICEP2. All of the models considered predict negligible running of the scalar spectral index, and would be ruled out by a detection of running. 
\end{abstract}

\maketitle


\begin{figure}
 \insertfig{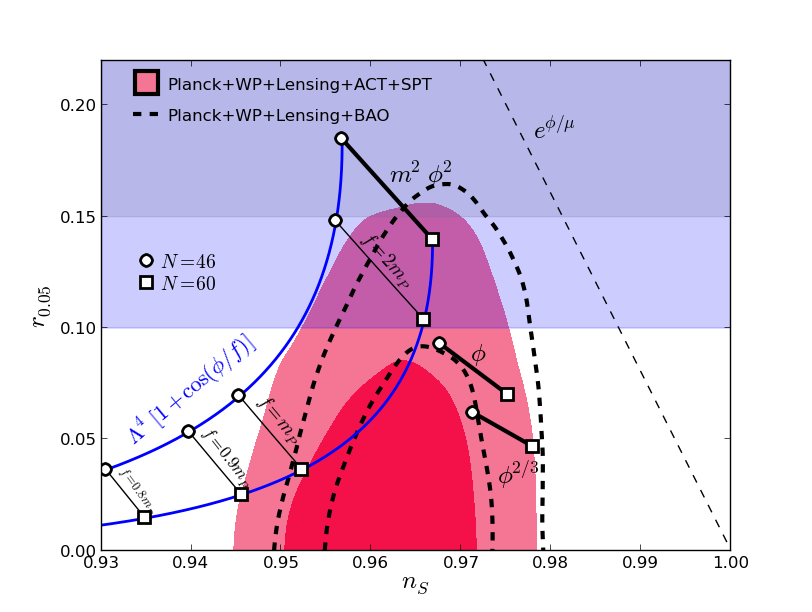}
  \caption[Cosine Natural Inflation]{
    {\bf Original Natural Inflation (Cosine Potential)}:  
   The band between the (solid/blue) lines running from approximately the lower left
    up to the middle of the plot are predictions for Natural Inflation for constant $N$ and varying $f$,
    where $N$ is the number of e-foldings prior to the end of inflation
    at which current modes of scale $k = 0.002$ Mpc$^{-1}$
    were generated and $f$ is the width of the potential.
    The  range of values of $N$ reflect uncertainties in reheating after inflation
    as described in the text. 
   Filled red (nearly vertical) regions are the
    parameter spaces allowed by Planck plus other CMB data as indicated at 68\% and 95\% C.L.'s.  Dotted regions are the parameter spaces allowed by
    Planck + WMAP Polarization + Lensing +BAO at 68\% and 95\% C.L.'s.  Horizontal blue bands correspond to 1(2) $\sigma$ measurements
    of $r$ from BICEP2 for the case of no running.  The predictions match the data for trans-Planckian $f$. Also shown are the axion monodromy potential $V \propto \phi^{2/3}$ and the linear potential $V \propto \phi$, which are inconsistent with the BICEP2 limit at $2\sigma$, and the power-law inflationary potential $V \propto \exp{\left(\phi/\mu\right)}$ .}
  \label{fig:cosine}
\end{figure}

\section{\label{sec:Intro} Introduction}
In a paper published in 1981, Guth proposed inflation \cite{Guth:1980zm} to solve several
cosmological puzzles: an early period of accelerated expansion explains
the homogeneity, isotropy, and flatness of the universe, as well as the
lack of relic monopoles.  Subsequently, Linde \cite{Linde:1981mu}, as well as Albrecht and Steinhardt \cite{Albrecht:1982wi}
suggested rolling scalar fields
as a mechanism to drive the dynamics of inflation (see 
\cite{Kazanas:1980tx,Starobinsky:1980te,Sato:1981ds,Sato:1980yn,Mukhanov:1981xt,Mukhanov:2003xw,Linde:1983gd} for important
early work).  While inflation results in an approximately
homogeneous universe, inflation models also predict small
inhomogeneities.  Observations of inhomogeneities via the cosmic
microwave background (CMB) anisotropies and structure formation 
provide strong tests of inflation models. 

In 1990, Freese, Frieman, and Olinto proposed the paradigm of natural inflation  \cite{Freese:1990rb} to 
solve theoretical problems of rolling inflation models. 
Most inflation models suffer from a potential drawback: to match
various observational constraints, namely CMB anisotropy measurements and the requirement of sufficient inflation,
the height of the inflaton potential must be of a much smaller scale
than that of the width, by many orders of magnitude (\ie, the potential
must be very flat).  
This requirement of two very different mass scales
is what is  known as the ``fine-tuning'' problem in inflation, since
very precise couplings are required in the theory to prevent radiative
corrections from bringing the two mass scales back to the same level.
The natural inflation model (NI) uses shift symmetries to generate a
flat potential, protected from radiative corrections, in a natural way
\cite{Freese:1990rb}. 

Natural inflation models use ``axions'' as the inflaton, the 
field responsible for inflation, where the term ``axion'' is used loosely for a field which has a  
 flat potential as a result of a shift symmetry, i.e. the potential
is unchanged under the transformation $\phi \rightarrow \phi + {\rm constant}$.  During the
early Universe, the inflaton field rolls along this flat potential for a long time, giving
rise to the long period of inflationary expansion that solves the cosmological problems
described above. Of course the shift symmetry must eventually be broken to allow the inflaton to
roll to a minimum of the potential and inflationary expansion to proceed and finally stop. In this sense
the inflaton in NI is an ``axion,'' or a ``pseudo-Nambu-Goldstone boson,'' with a nearly flat potential, exactly as
required by inflation.

In the original natural inflation model proposed in 1990, the inflaton was directly 
modeled after the QCD axion, though with different mass scales.  In this original model,
the shape of the potential is a cosine, exactly as for the QCD axion.  
To match CMB observations, the height of the potential in the original Cosine NI is required
to be $\sim 10^{16}$ GeV while the width is required to 
be $\gae 10^{19}$ GeV, as we will see in detail later in the paper.  
In 1995, WHK and K.T.\ Mahanthappa considered NI
potentials generated by radiative corrections in models with explicitly
broken Abelian \cite{Kinney:1995xv} and non-abelian \cite{Kinney:1995cc}
symmetries.  We will call these models
KM Natural Inflation.  Since that time many other variants of natural inflation have been proposed.
A notable example is axion monodromy, where an axion arising in string compactification is the inflaton;
the potential in this case is not periodic and instead can be linear or increasing as $\phi^{2/3}$.  Remarkably,
the data are now of sufficient accuracy to differentiate between these different types of NI models.

Over the past decade Cosmic Microwave Background (CMB) observations
have confirmed basic predictions of inflation and are in addition providing stringent tests 
of individual inflationary models.
  First, generic predictions of inflation
match the observations: the universe has a critical density
($\Omega=1$), the density perturbation spectrum is nearly scale
invariant, and superhorizon fluctuations are evident. Second, current
data  differentiate between inflationary models and
rule some of them out \cite{Spergel:2006hy,Alabidi:2006qa,Peiris:2006ug,Easther:2006tv,Seljak:2006bg,Kinney:2006qm,Martin:2006rs,Martin:2013tda,Martin:2013nzq,Peiris:2006sj}. For example, quartic potentials  and generic tree-level
hybrid models were disfavored already by  WMAP data.  The Planck satellite data has produced
powerful tests of single-field rolling models \cite{Ade:2013uln}.  It has
placed strong bounds on non-Gaussianity of the data, ruling out many non-minimal models including
variants with multiple fields, non-canonical kinetic terms, and non-Bunch-Davies vacua \cite{Ade:2013ydc}. 
To quote the Planck team,
``With these results, the paradigm of standard single-field inflation has survived its most stringent tests to date.''

Most recently, BICEP2 has made a ground-breaking discovery \cite{Ade:2014xna}. In addition to density perturbations, quantum fluctuations in inflation should
produce gravitational waves that would appear as B-modes in polarization data.  BICEP2 reported the first discovery of these gravity waves.
They find that the observed B-mode power spectrum is well-
fit by a lensed $\Lambda$-CDM + tensor theoretical model with tensor/scalar ratio $r = 0.20^{+0.07}_{-0.05}$, with the null hypothesis disfavored at 
7.0 $\sigma$.  Alternatively, if running of the spectral index is allowed, the combined Planck and BICEP data could have a different best fit.  
In this paper we restrict our studies to the case of no running, consistent with the predictions of the simplest Natural Inflation models. 
Thus we will take as our lower bound on $r$:
\begin{equation}
\label{eq:lowerbound}
r > 0.15\ {\mathrm{at}}\ 1 \sigma\ {\mathrm{and}}\ r>0.1\ {\mathrm{at}}\ 2 \sigma. 
\end{equation}
It is the purpose of this paper to test natural inflation models
with the Planck and BICEP2 data.

For comparison with the approach of taking the lower bound on $r$ from BICPE2, 
we also perform a joint likelihood analysis of Planck and BICEP2 (including  data from some other experiments as described below).
We compare the predictions of  the Cosine NI model with the 68\% and 95\% confidence level regions of this joint likelihood analysis.

Inflation models predict two types of perturbations, scalar and tensor,
which result in density and gravitational wave fluctuations,
respectively.  Each is typically characterized by a fluctuation
amplitude ($\Pscalarrt$ for scalar and $\Ptensorrt$ for tensor, with
the latter usually given in terms of the ratio $r \equiv
\Ptensor/\Pscalar$) and a spectral index ($\ns$ for scalar and
$\nt$ for tensor) describing the mild scale dependence of the
fluctuation amplitude.  The amplitude $\Pscalarrt$ is normalized
by the height of the inflationary potential.  The inflationary
consistency condition $r = -8 \nt$ further reduces the number of free
parameters to two, leaving experimental limits on $\ns$ and $r$ as the
primary means of distinguishing among inflation models.  Hence,
predictions of models are presented as plots in the $r$-$\ns$ plane.

The amplitude of the gravity waves and hence the value of $r$ is determined by the height of the potential,
i.e., the energy scale of inflation.   The relationship is given by
\begin{equation}
V  = (2.2 \times 10^{16} {\rm GeV})^4 \frac{r}{0.2} .
\end{equation}
Thus the BICEP2 bound $r>0.1$ at 2$\sigma$ requires the height of the potential to be at least $10^{16}$ GeV.
Inflation is probing the GUT scale.   Further, the width of the potential must exceed the
 the well-known Lyth Bound for single-field inflation \cite{Lyth:1996im}, which relates the tensor/scalar ratio to the 
field excursion $\Delta \phi$ during inflation, 
\begin{equation}
\Delta\phi \geq \mpl \sqrt{\frac{r}{4 \pi}}. 
\end{equation}
With $r \sim 0.2$, inflation potentially becomes an interesting test of physics beyond the Planck scale. 

The major results of this paper can be seen in Figures 1-5.   
The predictions of natural inflation models are plotted in the $r$-$\ns$ plane and compared
to data from the Planck and BICEP2/Keck data.  The predictions are plotted
 for various parameters: the width $f$ of the potential and number of e-foldings $N$
before the end of inflation at which present day fluctuation modes of
scale $k=0.002$ Mpc$^{-1}$ were produced.  $N$ depends upon the
post-inflationary universe and is $\sim 46-60$.  Also shown in the
figure are the observational constraints from Planck and BICEP2.  Figures 1-4 apply the lower bound on
$r$ in Eqn(\ref{eq:lowerbound}).
Figure 1 shows the original
Cosine NI model; Figure 2 the KM NI model, and Figure 3 summarizes 
a variety of potentials.  Figure 4 shows a Higgs potential for comparison.  In Figure 5 (for comparison with Figure 1), 
we plot the predictions of  the Cosine NI model vs. the 68\% and 95\% confidence level regions of the joint likelihood analysis of Planck and BICEP2 data.
Our primary result is that the original Natural Inflation Model and KM NI  are consistent with current observational
constraints.

In this paper we take $\mpl = 1.22 \times 10^{19}$ GeV.  Our result
extends upon previous analyses of NI \cite{Freese:2004un} and \cite{Savage:2006tr}
that was based upon WMAP's first year data \cite{Spergel:2003cb} and third year data.
Even earlier analyses \cite{Adams:1992bn,Moroi:2000jr}  placed
observational constraints on this model using COBE data
\cite{Smoot:1992td}.  Other papers have studied inflation models (including NI)  in light of the WMAP1 and WMAP3 data
\cite{Alabidi:2006qa,Alabidi:2005qi} and in light of Planck data \cite{Tsujikawa:2013ila}.

In  previous papers  \cite{Savage:2006tr} \cite{Freese:2008if}, we found
how far down the potential the field is at the time structure is
produced, and found that for $f \gg \mpl$ the relevant part of the
potential is indistinguishable from a quadratic potential (yet has the advantage that the required flatness is well- motivated). 
 Indeed one can see that $V\sim m^2 \phi^2$ matches
all the data. We will examine one other model with a GUT-scale Higgs-like potential, and show that it too can match the data.
The BICEP2 data have substantially reduced the number of inflationary models that agree with data.

We will begin by discussing the model of natural inflation in
\refsec{NI}: the motivation, the potential, and relating pre- and post-inflation scales. We will describe which of the
natural inflation models we plan to compare to data.  In
 \refsec{Fluctuations}, we will examine the scalar and tensor
perturbations predicted by NI models and compare them with Planck and BICEP2
data in \refsec{Results}.  We conclude in \refsec{Conclusion}.

\section{The Model of Natural Inflation\label{sec:NI}}

\subsection{\label{sec:Motivation} Motivation}

To satisfy a combination of constraints on inflationary models, in
particular, sufficient inflation and microwave background anisotropy
measurements \cite{Spergel:2003cb,Spergel:2006hy}, the
potential for the inflaton field must be very flat.  For a general
class of inflation models involving a single slowly-rolling field, it
has been shown that the ratio of the height to the (width)$^4$ of the
potential must satisfy \cite{Adams:1990pn}
\begin{equation} \label{eqn:Vratio}
  \chi \equiv \Delta V/(\Delta \phi)^4 \le {\cal O}(10^{-6} - 10^{-8})
  \, , 
\end{equation}
where $\Delta V$ is the change in the potential $V(\phi)$ and $\Delta
\phi$ is the change in the field $\phi$ during the slowly rolling
portion of the inflationary epoch.  Thus, the inflaton must be
extremely weakly self-coupled, with effective quartic self-coupling
constant $\lambda_{\phi} < \orderof{\chi}$ (in realistic models,
$\lambda_{\phi} < 10^{-12}$).  The small ratio of mass scales required
by \refeqn{Vratio} quantifies how flat the inflaton potential must be
and is known as the ``fine-tuning'' problem in inflation.
Reviews of inflation can be found in Ref.~\cite{Bassett:2005xm,Kinney:2009vz,Baumann:2009ds}.

Three approaches have been taken toward this required flat potential
characterized by a small ratio of mass scales.  First, some simply say
that there are many as yet unexplained hierarchies in physics, and
inflation requires another one.  The hope is that all these
hierarchies will someday be explained.  In these cases, the tiny
coupling $\lambda_{\phi}$ is simply postulated \textit{ad hoc} at tree
level, and then must be fine-tuned to remain small in the presence of
radiative corrections.  But this merely replaces a cosmological
naturalness problem with unnatural particle physics.  Second, models
have been attempted where the smallness of $\lambda_{\phi}$ is
protected by  supersymmetry.  Even if such a model succeeded (most suffer 
from the famous $\eta$-problem), the required mass hierarchy, while stable, is
itself unexplained.  It would be preferable if such a hierarchy, and thus inflation itself,
arose dynamically in particle physics models.

Hence, in 1990 a third approach was proposed, Natural Inflation
\cite{Freese:1990rb}, in which the inflaton potential is flat due to
shift symmetries.  The original model followed the physics of the QCD axion
though later variants have generalized.    The potential is exactly
flat due to a shift symmetry under $\phi \rightarrow \phi + \textrm{
constant}$. As long as the shift symmetry is exact, the inflaton
cannot roll and drive inflation, and hence there must be additional
explicit symmetry breaking.  Then these particles become pseudo-Nambu
Goldstone bosons (PNGBs), with nearly flat potentials, exactly as
required by inflation.  The small ratio of mass scales required by
\refeqn{Vratio} can easily be accommodated. For example, in the case
of the QCD axion, this ratio is of order $10^{-64}$.  While inflation
clearly requires different mass scales than the axion, the point is
that the physics of PNGBs can easily accommodate the required small
numbers.

The NI model was first proposed and a simple analysis performed in
\cite{Freese:1990rb}.  Then, in 1993, a second paper followed which
provides a much more detailed study \cite{Adams:1992bn}.  
Many types of candidates have subsequently been explored for natural
inflation.  For example, WHK and K.T.\ Mahanthappa considered NI
potentials generated by radiative corrections in models with explicitly
broken Abelian \cite{Kinney:1995xv} and non-abelian \cite{Kinney:1995cc}
symmetries.  We will mention a few others.   

We will see that cosine NI requires the width of the
potential to be trans-Planckian.  Such a scenario is difficult to accommodate in
string theory.  Thus many authors have proposed other variants of NI, taking advantage of the shift
symmetry offered by "axions," and looking for extensions
of the original cosine potential that accommodate smaller values of $f$.   Kim,
Nilles \& Peloso \cite{Kim:2004rp} as well as the idea of N-flation
\cite{Dimopoulos:2005ac,Grimm:2007hs} generalized the original NI model to include
two or more axions, and showed that an \textit{effective} potential of
$f \gg \mpl$ can be generated from multiple axions, each with
sub-Planckian scales.
Ref.~\cite{Kawasaki:2000yn} used shift symmetries
in Kahler potentials to obtain a flat potential and drive natural
chaotic inflation in supergravity.  Additionally,
\cite{ArkaniHamed:2003wu,ArkaniHamed:2003mz} examined natural
inflation in the context of extra dimensions and \cite{Kaplan:2003aj}
used PNGBs from little Higgs models to drive hybrid inflation.  Also,
\cite{Firouzjahi:2003zy,Hsu:2004hi} use the natural inflation idea of
PNGBs in the context of braneworld scenarios to drive inflation.
Freese \cite{Freese:1994fp} suggested using a PNGB as the rolling
field in double field inflation \cite{Adams:1990ds} (in which the
inflaton is a tunneling field whose nucleation rate is controlled by
its coupling to a rolling field).  Ref. \cite{Kaloper:2008fb,Kaloper:2011jz} found
 a quadratic potential  in theories where an "axion" field mixes with a 4-form.
Ref. \cite{Germani:2010hd,Germani:2011ua} used coupling of 
the inflaton kinetic term to the Einstein tensor to allow NI with $f << m_{pl}$ by enhancing the gravitational friction
 acting on the inflaton during inflation.   Ref. \cite{Czerny:2014wza,Czerny:2014xja} suggested a
 "multi-natural" inflation model in which the single-field inflaton potential consists of two or more 
 sinusoidal potentials  with a possible non-zero relative phase
 (such as may arise if a complex scalar field  is coupled to two sets of quark and anti- quark fields).
 We will focus in this paper on 
single field implementations of NI.

\subsection{\label{sec:Potential} Potential}

We present a variety of natural inflation potentials. The feature they all share is a shift symmetry
that maintains the required flatness of the potential.

\subsubsection{Original Natural Inflation (Cosine Potential)}
In the original natural inflation model, modeled after the QCD axion,
 the PNGB potential resulting from explicit breaking of a shift symmetry is  of
the form
\begin{equation} \label{eqn:potential}
  V(\phi) = \Lambda^4 [1 \pm \cos(N\phi/f)] \, .
\end{equation}
We will take the positive sign in \refeqn{potential} (this choice has
no effect on our results) and take $N = 1$, so the potential, of
height $2 \Lambda^4$, has a unique minimum at $\phi = \pi f$ (the
periodicity of $\phi$ is $2 \pi f$).

For appropriately chosen values of the mass scales, \eg\ $f \gae \mpl$
and $\Lambda \sim \mgut \sim 10^{16}$ GeV, the PNGB field $\phi$ can
drive inflation.  This choice of parameters indeed produces the small
ratio of scales required by \refeqn{Vratio}, with $\chi \sim
(\Lambda/f)^4 \sim 10^{-13}$.  For  $\Lambda \sim
10^{15}$-$10^{16}$~GeV we have $f \sim \mpl$, yielding an inflaton mass
$m_\phi = \Lambda^2/f \sim 10^{11}$-$10^{13}$~GeV.    For $f \gg \mpl$, the inflaton
becomes independent of the scale $f$ and is $m_\phi \sim 10^{13}$ GeV.

\subsubsection{Low-scale inflation}

It is possible that the shift symmetry responsible for providing the stability of the mass hierarchy in NI is respected to such a degree that the mass term for the inflaton is identically zero. In such a case, an effective expansion for the inflaton potential can be written
\begin{equation}
V\left(\phi\right) = V_0 - \sum_p{\lambda_p \left(\frac{\phi}{\mu}\right)^p},
\end{equation}
where the leading order operator for $\phi \ll \mu$ is of order $p > 2$. In this case, a remarkable cancellation occurs, such that the spectral index is independent of the mass scales in the potential, and depends {\it only} on the number of e-folds of inflation \cite{Kinney:1995cc},
\begin{equation}
\ns - 1 = - \left(\frac{2}{N}\right) \frac{p - 1}{p - 2},
\end{equation}
where $N \sim 60$ is the number of e-folds before the end of inflation. For $N = [46,60]$, consistent with a reheat temperature above $1\ {\rm TeV}$, and assuming an even exponent $p \geq 4$, the scalar spectral index is confined to a narrow range, $\ns = [0.935,0.967]$, which overlaps nicely with the region favored by Planck. Because the spectral index is independent of the scales in the potential, such models can successfully generate inflation for $\mu \ll m_{\rm Pl}$. \reffig{lowscale} shows the predictions of various low-scale models relative to Planck and BICEP. Such models with $\mu < m_{\rm Pl}$ are strongly ruled out by the tensor mode detection of BICEP. 

In 1995, Kinney and Mahanthappa proposed a realization of such low-scale inflation scenarios in which the inflaton potential is generated by radiative corrections in an explicitly broken SO(3) gauge symmetry. In this scenario, the inflaton is a pseudo-Goldstone mode with potential
\begin{equation}
V = V_0  \left\lbrace \sin^4\left(\frac{\phi}{\mu}\right) \log\left[g^2 \sin^2\left(\frac{\phi}{\mu}\right)\right] - \log\left[g^2\right]\right\rbrace.
\end{equation}
In the low-scale limit $\mu \ll \mpl$, such potentials reduce to the quartic hilltop type at leading order in a Taylor expansion,
\begin{equation}
V \simeq V_0 - \lambda \left(\frac{\phi}{\mu}\right)^4 + \cdots
\end{equation}
The low-scale limit of these models is inconsistent with the data, but extension of the model to large field values, $\mu > m_{\rm Pl}$ is possible, and is still consistent with the BICEP2 data. \reffig{KM} shows this model relative to the Planck and BICEP2 constraints. 

\subsubsection{String-motivated Axion Potentials}

Axion monodromy \cite{Silverstein:2008sg} is a shift-symmetric string-motivated version of Natural Inflation which evades the super-Planck scale width of the cosine potential by analytic continuation on a compact manifold, resulting in an effective field range larger than $\mpl$.  A resulting potential $V \propto \phi^{2/3}$ is inconsistent with the BICEP2 data at $2\sigma$.  Linear Axion Monodromy \cite{McAllister:2008hb} with $V \propto \phi$  is also inconsistent with BICEP2 to $2\sigma$. Versions of axion monodromy with additional couplings to heavy degrees of freedom can produce larger tensor amplitudes \cite{Dong:2010in}, consistent with BICEP2. 

\begin{figure}
\end{figure}

\begin{figure}
  \insertfig{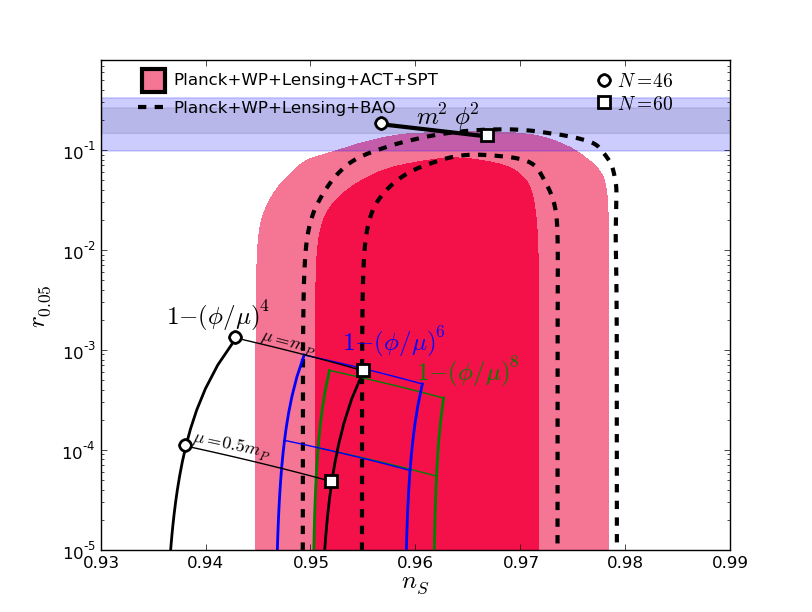}
  \caption[Low-scale inflation]{
    {\bf Low-scale models of natural inflation} Low-scale models with $\mu < m_{\rm Pl}$ are shown relative to the constraints from Planck and BICEP2. These models are strongly ruled out by the BICEP2 detection of tensor modes.  Labels same as in Figure 1 (roughly, solid lines are theoretical predictions; red is Planck data; blue is BICEP2 data). }
  \label{fig:lowscale}
\end{figure}

\begin{figure}
  \insertfig{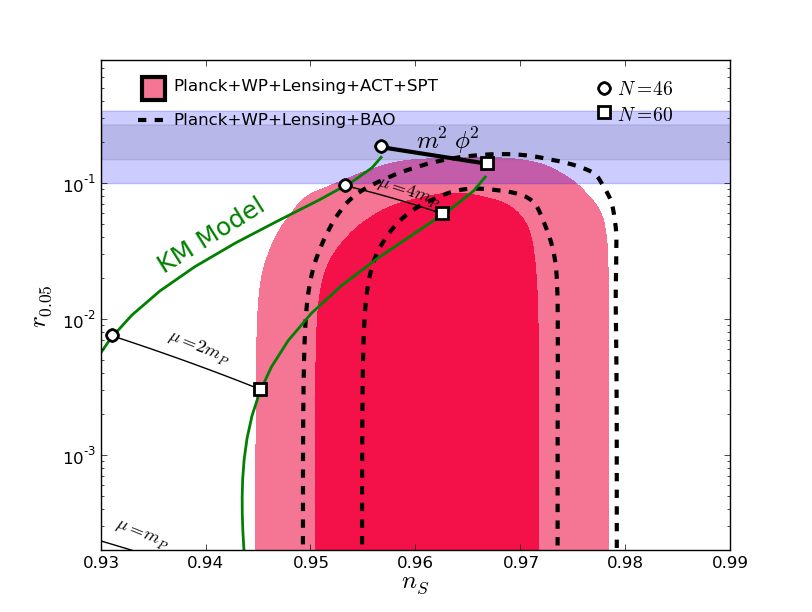}
  \caption[KM natural inflation]{
    {\bf KM Model of natural inflation}   Labels same as in Figure 1 (roughly, solid lines are theoretical predictions; red is Planck data; blue is BICEP2 data). }
  \label{fig:KM}
\end{figure}

\subsection{\label{sec:Horizon} Relating Pre- and Post-Inflation Scales}

To test inflationary theories, present day observations must be
related to the evolution of the inflaton field during the inflationary
epoch.  Here we show how a comoving scale $k$ today can be related
back to a point during inflation.  We need to find the value of $N_k$,
the number of e-foldings before the end of inflation, at which
structures on scale $k$ were produced.

Under a standard post-inflation cosmology, once inflation ends, the
universe undergoes a period of reheating. Reheating can be
instantaneous or last for a prolonged period of matter-dominated
expansion.  Then reheating ends at $T <T_\textrm{RH}$, and the
universe enters its usual radiation-dominated and subsequent
matter-dominated history.  Instantaneous reheating ($\rhoRH = \rho_e$)
gives the minimum number of e-folds as one looks backwards
to the time of perturbation production, while a prolonged period of
reheating gives a larger number of e-folds.

The relationship between scale $k$ and the number of e-folds $N_k$
before the end of inflation has been shown to be \cite{Lidsey:1995np}
\begin{equation} \label{eqn:Nk}
  N_k = 62 - \ln\frac{k}{a_0 H_0}
            - \ln\frac{10^{16}\,\textrm{GeV}}{V_k^{1/4}}
            + \ln\frac{V_k^{1/4}}{\Ve^{1/4}}
            - \frac{1}{3} \ln\frac{\Ve^{1/4}}{\rhoRH^{1/4}} \, .
\end{equation}
Here, $V_k$ is the potential when $k$ leaves the horizon during
inflation, $\Ve = V(\phie)$ is the potential at the end of inflation,
and $\rhoRH$ is the energy density at the end of the reheat period.
Nucleosynthesis generally requires $\rhoRH \gae (\textrm{1 GeV})^4$,
while necessarily $\rhoRH \le \Ve$.  Since $\Ve$ may be of order
$\mgut \sim 10^{16}$ GeV or even larger, there is a broad
allowed range of $\rhoRH$; this uncertainty in $\rhoRH$ translates
into an uncertainty of 10 e-folds in the value of $N_k$ that
corresponds to any particular scale of measurement today.

Henceforth we will use $N$ to refer to the number of e-foldings prior
to the end of inflation that correspond to scale\footnote{The current horizon scale corresponds to
  $k \approx 0.00033$ Mpc$^{-1}$.  The difference in these two scales
  corresponds to only a small difference in e-foldings of
  $\Delta N \lae 2$: while we shall present parameters evaluated at
  $k = 0.05$ Mpc$^{-1}$, those parameters evaluated at the current
  horizon scale will have essentially the same values (at the few
  percent level).
  } $k = 0.002$
Mpc$^{-1}$.
Under the standard cosmology\footnote{However, if one were to consider
non-standard cosmologies \cite{Liddle:2003as}, the range of possible
$N$ would be broader. }, this scale corresponds to
$N\sim$46-60 (smaller $N$ corresponds to smaller $\rhoRH$), with a
slight dependence on $f$.

\section{\label{sec:Fluctuations} Perturbations}

As the inflaton rolls down the potential, quantum fluctuations lead
to metric perturbations that are rapidly inflated beyond the horizon.
These fluctuations are frozen until they re-enter the horizon during
the post-inflationary epoch, where they leave their imprint on large
scale structure formation and the cosmic microwave background (CMB)
anisotropy \cite{Guth:1982ec,Hawking:1982cz,Starobinsky:1982ee}.

\subsection{\label{sec:Scalar} Scalar (Density) Fluctuations}

The perturbation amplitude for the density fluctuations (scalar modes) 
produced during inflation is given by
\cite{Mukhanov:1985rz,Mukhanov:1988jd,Mukhanov:1990me,Stewart:1993bc}
\begin{equation} \label{eqn:Pscalar}
   \Pscalarrt(k) = \frac{H^2}{2\pi\dot{\phi}_k} \, .
\end{equation}
Here, $\Pscalarrt(k) \sim \frac{\delta\rho}{\rho}|_\textrm{hor}$ 
denotes the perturbation amplitude when a given wavelength re-enters the
Hubble radius in the radiation- or matter-dominated era, and
the right hand side of \refeqn{Pscalar} is to be evaluated when
the same comoving wavelength ($2\pi/k$) crosses outside the horizon
during inflation.

Normalizing to the COBE \cite{Smoot:1992td} or WMAP
\cite{Spergel:2006hy} an\-iso\-tropy measurements gives $\Pscalarrt \sim
10^{-5}$.  This normalization can be used to approximately fix the
height  $\Lambda$ of the potential \refeqn{potential}.  The largest
amplitude perturbations on observable scales are those produced
$N \sim 60$ e-folds before the end of inflation (corresponding to the
horizon scale today), when the field value is $\phi = \phi_N$.  For cosine Natural Inflation, under
the SR approximation, the amplitude on this scale takes the value
\begin{equation} \label{eqn:Pscalar2}
  \Pscalar \approx \frac{128\pi}{3}
                   \left( \frac{\Lambda}{\mpl} \right)^4
                   \left( \frac{f}{\mpl} \right)^2
                   \frac{[1 + \cos(\phi_N/f)]^3}{\sin^2(\phi_N/f)} \, .
\end{equation}

The fluctuation amplitudes are, in general, scale dependent.  The
spectrum of fluctuations is characterized by the spectral index $\ns$,
\begin{equation} \label{eqn:ns}
  \ns - 1 \equiv  \frac{\rmd\ln\Pscalar}{\rmd\ln k}
          \approx -\frac{1}{8\pi} \left( \frac{\mpl}{f} \right)^2
                  \frac{3 - \cos(\phi/f)}{1 + \cos(\phi/f)} \, .
\end{equation}
  For
small $f$, $\ns$ is essentially independent of $N$, while for
$f \gae 2\mpl$, $\ns$ has essentially no $f$ dependence.  Analytical
estimates can be obtained in these two regimes:
\begin{equation} \label{eqn:nsA}
  \ns \approx
  \begin{cases}
    1 - \frac{\mpl^2}{8 \pi f^2} \, ,
      & \textrm{for} \,\, f \lae \frac{3}{4}\mpl \\
    1 - \frac{2}{N} \, ,
      & \textrm{for} \,\, f \gae 2\mpl \, .
  \end{cases}
\end{equation}

  For $f \gae m_{pl}$, natural inflation predicts a small, $O(10^{-3}$), negative spectral index running\footnote{See Figure 6 in \cite{Savage:2006tr}}. 
    This small running is a prediction of the model.
  
  The Planck results as shown in Figure 1 led to the constraint
$f \gae 0.8\mpl$ at 95\% C.L.  With the inclusion of BICEP2 data,
the bound becomes stronger and $f$ must be trans-Planckian.

\begin{figure}
  \insertfig{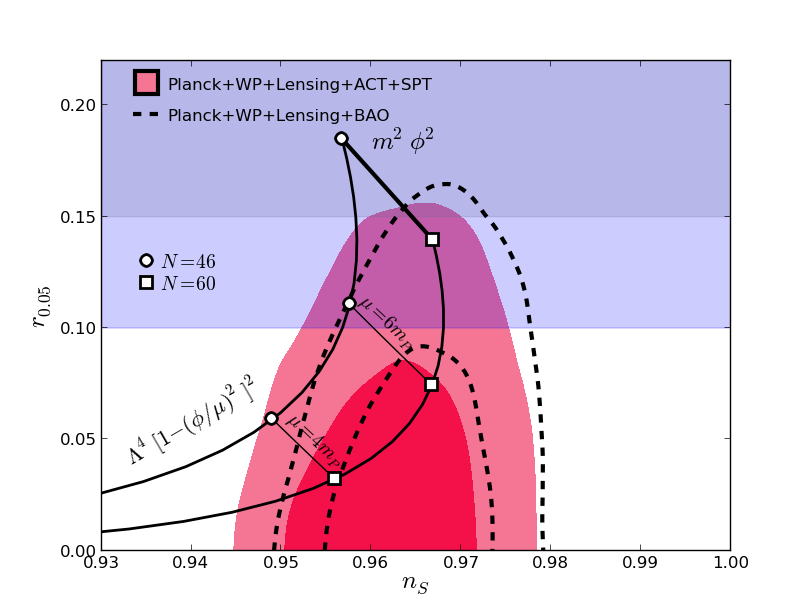}
  \caption[Higgs-like inflation]{
   {\bf  Higgs-like inflation}. Labels same as in Figure 1 (roughly, solid lines are theoretical predictions; red is Planck data; blue is BICEP2 data).
    }
  \label{fig:Higgs}
\end{figure}

\subsection{\label{sec:Tensor} Tensor (Gravitational Wave) Fluctuations}

In addition to scalar (density) perturbations, inflation also produces
tensor (gravitational wave) perturbations with amplitude
\begin{equation} \label{eqn:Ptensor}
  \Ptensorrt(k) = \frac{4H}{\sqrt{\pi}\mpl} \, .
\end{equation}
As mentioned in the introduction, the amplitude of the gravity waves is
directly proportional to the Hubble parameter and therefore is determined by the 
energy scale of the height of the potential.

Conventionally, the tensor amplitude is given in terms of the
tensor/scalar ratio
\begin{equation} \label{eqn:Pratio}
  r \equiv \Pratio = 16 \epsilon \, ,
\end{equation}
 For small $f$,
$r$ rapidly becomes negligible, while $f \to \frac{8}{N}$ for
$f \gg \mpl$.  

As mentioned in the introduction, 
in principle, there are four parameters describing scalar and tensor
fluctuations: the amplitude and spectra of both components, with the
latter characterized by the spectral indices $\ns$ and $\nt$
(we are ignoring any running here).  The amplitude of the scalar
perturbations is normalized by the height of the potential (the energy
density $\Lambda^4$).  The tensor spectral index $\nt$ is not
an independent parameter since it is related to the tensor/scalar ratio
$r$ by the inflationary consistency condition $r = -8 \nt$.
The remaining free parameters are the spectral index $\ns$ of the scalar
density fluctuations, and the tensor amplitude (given by $r$).  
Hence, a useful parameter space for plotting the model predictions
versus observational constraints is on the $r$-$\ns$ plane
\cite{Dodelson:1997hr,Kinney:1998md}.  

The Planck constraints are generated using the COSMOMC Markov Chain Monte Carlo code \cite{Lewis:2002ah}, marginalizing over a seven-parameter data set with flat priors: 
\begin{itemize}
\item{Dark Matter density $\Omega_{\rm M} h^2$.}
\item{Baryon density $\Omega_{\rm b} h^2$.}
\item{Reionization optical depth $\tau$.}
\item{The angular size  $\theta$ of the sound horizon at decoupling.}
\item{Scalar spectrum normalization $A_{\rm S}$.}
\item{Tensor/scalar ratio $r$.}
\item{Scalar spectral index $\ns$.}
\end{itemize}
The fit assumes a flat universe $\Omega_{\rm b} + \Omega_{\rm M} + \Omega_{\rm \Lambda} = 1$, with Cosmological Constant Dark Energy, $\rho_{\Lambda} = {\rm const.}$ Convergence is determined via a Gelman and Rubin statistic. Auxiliary data sets used are WMAP polarization (WP), in combination with the Atacama Cosmology Telescope (ACT) / South Pole Telescope (SPT) CMB measurements (solid contours in figures), and Baryon Acoustic Oscillation (BAO) data from Sloan Digital Sky Survey Data Release 9 \cite{Ahn:2012fh}, the 6dF Galaxy Survey \cite{Jones:2009yz}, and the WiggleZ Dark Energy Survey \cite{Blake:2011en} (dashed contours). 

\begin{figure}
  \insertfig{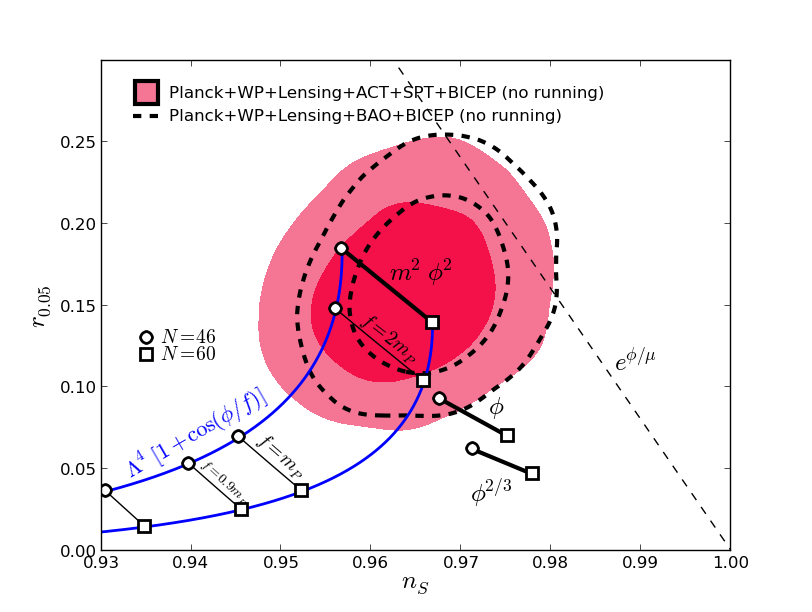}
  \caption[Joint Likelihood]{
   {\bf Cosine Natural Inflation: Joint likelihood} for Planck + WMAP Polarization + Lensing + BICEP, including ACT/SPT and BAO, for comparison with \reffig{cosine}. Inner contours are 68 \% confidence limits, outer contours are 95 \% confidence. Note that the linear potential $V \propto \phi$ is marginally consistent with the joint fit at 95\% confidence, as is the power-law potential $V \propto e^{\phi/\mu}$. 
    }
  \label{fig:joint}
\end{figure}

The BICEP2 allowed region is taken from Ref. \cite{Ade:2014xna} as $r = 0.2^{+0.07}_{-0.05}$. We plot this as a one-sigma (dark-) and two-sigma (light-) shaded region, separately from the Planck contours, which illustrates two important points: First BICEP2 alone does not provide constraint on the spectral index of scalar perturbations, since it only measures the amplitude of the tensor signal. Second, there is clearly tension between the BICEP2 constraint on the tensor/scalar ratio $r$ and the Planck constraint with no running of the spectral index, which sets a 95\%-confidence upper limit of $r<0.17$ \cite{Ade:2013uln}. While Planck and BICEP2 are consistent at the $2\sigma$ confidence level, they are significantly discrepant at the $1\sigma$ level. For comparison, a joint likelihood for Planck and BICEP is shown in \reffig{joint}. The BICEP2 contours are sensitive to the details of foreground removal \cite{Ade:2014xna}, which may help lower the best-fit $r$ slightly, and at least partially resolve the tension with Planck. Other possibilities for resolving the tension are an extra neutrino species \cite{Giusarma:2014zza}, or running of the spectral index \cite{Ade:2014xna} (the latter would invalidate all of the models considered here). For the purposes of this paper, we take the quoted best-fit region for BICEP2 at face value, and investigate its implications.

\section{Results}
\label{sec:Results}
In \reffig{cosine}, we show the predictions of the original natural inflation model
together with the observational constraints.  Parameters
corresponding to fixed $N$= 46 and 60 with varying $f$ are shown
as (solid/blue) lines from the lower left to upper right.  The
orthogonal (black) lines correspond to fixed $f$ with varying $N$.
The  band between the blue lines are the values of $N$ consistent with standard
post-inflation cosmology for reheat temperatures above the
nucleosynthesis limit of $\sim$1 GeV, as discussed previously.  The
solid red regions are the allowed parameters at 68\% and 95\% C.L.'s from Planck + other CMB data.
The blue regions are the parameters at 1 and 2 $\sigma$ consistent with the BICEP2 discovery of B modes.
For a given $N$, a fixed point is reached for $f \gg \mpl$; that is,
$r$ and $\ns$ become essentially independent of $f$ for any
$f \gg \mpl$, and Natural Inflation becomes equivalent to a large-field $m^2 \phi^2$ model (note, however,
that in natural inflation this effectively power law potential is
produced via a natural mechanism). 
As seen in the figure, $f \lae \mpl$ is excluded.  However,
$f \gae \mpl$ falls well into the  allowed region and is thus
consistent with all data.    \reffig{cosine} also shows the axion monodromy potential with $V\sim \phi^{2/3}$, which is inconsistent with the BICEP2 constraint.
Linear potentials $V \propto \phi$ are inconsistent with the BICEP2 limit at the 95\% confidence level, but are marginally consistent with a joint Planck/BICEP2 limit at 95\%,
as is the power-law potential $V \propto e^{\phi/\mu}$.

In \reffig{KM} we show the predictions of KM inflation and compare them
to all data sets.  There is some tension in obtaining high enough values of $r$ in these models.  

In \reffig{lowscale} we plot the results for a variety of potentials for low-scale inflation models of the form 
$V(\phi) \propto 1- (\phi / \mu)^p$ with $p \gae 4$.   
These models would have had low scales; the energy scale
of the potentials as well as their widths could have been much lower than for the cosine model.  However, 
these low scales correspond to low values of $r$ and are ruled out by the BICEP2 data.

For comparison, in \reffig{Higgs} we study a Higgs-like potential, which has no connection to natural inflation. The potential is that of a Higgs-like particle at the GUT scale, with potential
\begin{equation}
V\left(\phi\right) = V_0 \left[1 - \left(\frac{\phi}{\mu}\right)^2\right]^2
\end{equation}
 One can see that
such a potential remains a good fit to the data as well, as long as the mass scales are high.

In Figure 5, we compare the predictions of the cosine Natural Inflation model vs. the joint likelihood for Planck + WMAP Polarization + Lensing + BICEP, including ACT/SPT and BAO, for comparison with \reffig{cosine}. Inner contours are 68\% confidence limits, outer contours are 95\% confidence.

\section{\label{sec:Conclusion} Conclusion}

Remarkable advances in cosmology have taken place in the past decade
thanks to Cosmic Microwave Background experiments.  The release of the
BICEP2 data is revolutionary and will lead 
to even more exciting times for inflationary cosmology.  The success of BICEP2 should motivate future missions even going to space.
Not only have generic
predictions of inflation been confirmed by a series of CMB experiments (though there are still outstanding
theoretical issues), but indeed individual inflation models are
being  tested with large classes already ruled out.

Currently the natural inflation model, which is
well-motivated on theoretical grounds of naturalness, is a good fit to
existing data.  In Figure 1, we showed that for the cosine potential with
 width $f \gae \mpl$ and height $\Lambda \sim \mgut$ the model is in good
agreement with all CMB data.  Natural inflation predicts very little
running, at the level of $10^{-3}$, and this will become a test of the model.
 Even for values $f\gg \mpl$ where the relevant parts of the
potential are indistinguishable from quadratic, natural inflation
provides a framework free of fine-tuning for the required potential.

Other than natural inflation, single-field models compatible with all existing data sets include the  $m^2 \phi^2$ quadratic potential (to which 
natural inflation asymptotes for large $f$ as mentioned above)
as well as the potential for a Higgs-like particle at the GUT scale (see Figure 4).  
The BICEP2 data have substantially reduced the number of inflationary models that agree with data.

 In summary, Natural Inflation represents a model which is both well-motivated and testable.
It is a  good fit to all cosmic microwave background (CMB) data and may be the correct description of an early inflationary expansion of the Universe.

\begin{acknowledgments}
   KF acknowledges the support of the DOE under grant DOE-FG02-95ER40899 and the Michigan
  Center for Theoretical Physics at the University of Michigan.
   WHK is supported in part by the National Science Foundation under
  grant NSF-PHY-1066278. WHK thanks Perimeter Institute for Theoretical Physics, where part of this work was completed, for hospitality. This work was performed in part at the University at Buffalo Center for Computational Research. 
  \end{acknowledgments}

\bibliography{natinflation}

\end{document}